\documentclass[journal]{IEEEtran}

\usepackage{amsmath, amssymb, amsthm}
\usepackage{hyperref}

\def\F{{\mathbb F}}
\def\Z{{\mathbb Z}}

\def\R{{\mathbb R}}
\newtheorem{thm}{Theorem}[section]
\newtheorem{prop}[thm]{Proposition}

\newtheorem{lem}[thm]{Lemma}

\theoremstyle{definition}

\newtheorem{rem}[thm]{Remark}
\newtheorem{ex}[thm]{Example}
\newenvironment{preuve}[1][Proof]{\noindent\textbf{#1.} }{\ \rule{0.5em}{0.5em}}

\usepackage{xcolor}

\usepackage{graphicx}
\graphicspath{{./}{fig/}}
\newcommand\transpose[1]{#1^\mathsf{T}} 

\def\rank{\operatorname{rank}}

\newcommand\HW[1]{w_H\left(#1\right)}
\def\independent{\perp\!\!\!\perp}

\begin{document}

\title{Higher-order CIS codes}

\author{%
Claude   Carlet    \hspace{0.5cm}
Finley   Freibert  \hspace{0.5cm}
Sylvain  Guilley   \hspace{0.5cm}
Michael  Kiermaier \hspace{0.5cm}
Jon-Lark Kim       \hspace{0.5cm}
Patrick  Sol\'{e}%
\thanks{%
C. Carlet is with LAGA, Universities of Paris 8 and Paris 13; CNRS, UMR 7539; Address: University of Paris 8, Department of Mathematics,
2 rue de la libert\'e, 93\,526 Saint-Denis cedex 02, France.
F. Freibert is with Department of Mathematics,
Ohio Dominican University,
Columbus, OH 43219, USA.
S. Guilley is with Institut Mines-T\'el\'ecom, T\'el\'ecom-ParisTech,
37/39 rue Dareau 75\,014 Paris, France and
Secure-IC S.A.S., 80 avenue des Buttes de Co\"esmes, 35\,700 Rennes, France.
M. Kiermaier is with the Department of Mathematics, Bayreuth University, 95440 Bayreuth, Germany.
J.-L. Kim is with Department of Mathematics,
Sogang University,
Seoul 121-742, South Korea.
J.-L. Kim was supported by Basic Research Program through the National Research Foundation of Korea (NRF) funded by the Ministry of Education (NRF-2013R1A1A2005172) and by the Sogang University Research Grant of 201210058.01.
P. Sol\'{e} is with CNRS/LTCI, UMR 5141, T\'el\'ecom-ParisTech,
46 rue Barrault 75\,634 Paris cedex 13, France and Math Dept of King Abdulaziz University, Jeddah, Saudi Arabia.}}

\date{}

\maketitle
\begin{abstract}
We introduce  {\bf complementary information set codes} of higher-order.
A binary linear code of length $tk$ and dimension $k$ is called a
complementary information set code of order $t$ ($t$-CIS code for short) if it has
$t$ pairwise disjoint information sets.
The duals of such codes permit to reduce the cost of masking cryptographic algorithms
against side-channel attacks.
As in the case of codes for error correction, given the length and the dimension of a $t$-CIS code, we look for the highest possible minimum distance.
In this paper, this new class of codes is investigated. The existence of good long CIS codes of order $3$ is derived by a counting argument.
General constructions based on cyclic and quasi-cyclic codes and on the building up construction are given.  A formula similar to a mass formula is given. A classification of 3-CIS codes of length $\le 12$ is given. Nonlinear codes better than linear codes are derived by taking binary images of $\Z_4$-codes.
A general algorithm based on Edmonds' basis packing algorithm from matroid theory is developed with the following property: given a binary linear code
of rate $1/t$ it either provides $t$ disjoint information sets or proves that the code is not $t$-CIS. Using this algorithm, all optimal or best known $[tk, k]$ codes where $t=3, 4, \dots, 256$ and $1 \le k \le \lfloor 256/t \rfloor$ are shown to be $t$-CIS for all such $k$ and $t$, except for $t=3$ with  $k=44$ and $t=4$ with $k=37$.

\end{abstract}

\section{Introduction}
In a recent paper~\cite{2CIS} was introduced the notion of {\bf complementary information set codes} (CIS codes for short), that is, binary linear codes of rate one half that admit a pair of complementary
information sets. In the present work we consider binary codes of rate $1/t$ that admit $t$ pairwise disjoint information sets, where $t \ge 2$. These we call
complementary information set codes of order $t$ ($t$-CIS codes for short). Like for \cite{2CIS} the motivation (developed below) comes from security of embedded cryptographic hardware;
in particular from the approach of \emph{leakage squeezing} to counter side-channel attacks by Boolean masking of cryptographic computations.

\subsection{Connection with the state-of-the-art, and new contributions}

There are two main differences between the present work and the paper~\cite{2CIS}.
First, there is no analogue of self-dual codes for $t>2$.
In particular, we have defined $t$-CIS codes as codes of rate $1/t$ (whose minimum distance is required to be as large as possible) for the sake of simplicity.
From the viewpoint of applications, however, a definition in terms of rate $(1-1/t)$ codes
of large dual distance
might have been more natural (see Sections~\ref{sec-motivation} and~\ref{sec-boolean_functions});
since all codes considered in this paper enjoy standard duals (when linear) or formal duals (when $\Z_4$-linear), no difficulty arises.
Indeed, all $\Z_4$-linear codes have formal duals~\cite{HKCSS}.
Next,
we have found a general algorithm
to test $t$-CISness, whatever the value of $t\geq 2$ is, based on Jack Edmonds' base partition algorithm from matroid theory~\cite{Edm} (note that in~\cite{2CIS}, there is not such an algorithm to test CISness).  The complexity of this algorithm is polynomial while a naive algorithm runs in an exponential time~(see Sec. VII and~\cite{K73}). Thus this algorithm can be efficiently used to determine $2$-CIS codes introduced in~\cite{2CIS}.
While the paper~\cite{Edm} is self-contained, the reader can find some more background on matroids in \cite{O,W}. For the reader acquainted with this theory, let it suffice to say that Edmonds' algorithm is applied to the matroid of linear dependence defined on the columns of the generator matrix of the code tested. This algorithm will allow us to show that there exist optimal or as good as best known $t$-CIS $[tk, k]$ codes where $t=3, 4, \dots, 256$ and $1 \le k \le \lfloor 256/t \rfloor$ except for $t=3$ with $k=44$ and $t=4$ with $k=37$, where $256$ is the upper limit on length for
 the best known binary linear codes in the Magma database~\cite{magma} and the Grassl tables~\cite{Grassl}.

For the common point with \cite{2CIS} one can cite the connection with vectorial Boolean functions, especially permutations of $\F_2^k$.
Indeed, the existence of $t$ disjoint information sets is equivalent to the existence of bijections.
By the definition of an information set, each codeword (of length $tk$) can be bijectively identified with the vector (say $x$) of length $k$ equal to its restriction to some information set $i$ ($0\leq i<t$).
Now, the restriction to the $i$-th information set ($0\leq i<t$) of the codeword is the image of $x$ by a bijection, denoted by $G_i$ in the paper.

Nonlinear permutations better than linear permutations are constructed in connection with $\Z_4$-codes with binary images better than the best or best known binary linear codes. A recently discovered  $\Z_4$-code of parameters $(24,4^6,18)$ is applied. The asymptotic properties
of long $3$-CIS codes of given rate $1/3$ are studied. They are shown to satisfy the Gilbert-Varshamov bound, and this fact cannot be deduced from known results on self-dual codes like for $t=2.$

\subsection{Application to masking schemes}

Masking schemes are methods to carry out cryptographic computations that resist (to some extent) side-channel attacks.
 Specifically, each sensitive variable (any intermediate variable that depends on the input or output and on a secret, e.g., a key) involved in the cryptographic algorithm is split in $t$ shares, obtained by randomizing $t-1$ vectors, called masks, of the same length as the sensitive variable, XORing all of them altogether with the sensitive variable to obtain a $t$-th share (this is standard masking) and
 processing, instead of the $t-1$ masks, their images by bijections (this so-called leakage squeezing method allows to make the attacks more difficult).
 The bijections which encode the masks are denoted by $F_i$ ($0 \leq i < t$) in the paper.
 A natural question is to quantify in which respect the proposed bijections improve the security.
 The problem is stated in Section~\ref{sec-motivation} and then solved in Section~\ref{sec-boolean_functions} (using $t$-CIS codes).
 When designing a masking scheme against side-channel attacks, a strategic decision is the choice for parameter $t$.


The parameter $t$ determines the number of shares the attacker shall collect to extract the key.
The way the shares are then combined to build a side-channel distinguisher is discussed in detail in Sec.~\ref{sec-motivation}.
When $t$ is larger, the countermeasure is more secure, but it is also more expensive:
the cost overhead varies
as $t^2$ in implementations not subject to glitches~\cite{RP} and
as $t^3$ otherwise (when the countermeasure must explicitly enforce the glitch-freeness~\cite{DBLP:conf/ches/ProuffR11}).
Implementations with $t=2$ are now state-of-the-art in commercial products, and $t \geq 3$ are recommended for forward-security applications.
Then, the application of leakage squeezing consists in making the exploitation of the residual leakage conveyed by the $t$ shares as chancy as possible.
The goal is to increase the order $d$ of the easiest high-order correlation attack~\cite{DBLP:conf/ctrsa/SchrammP06},
and by the same token to reduce the mutual information between the leakage and the sensitive variables~\cite{leakage}.
For \emph{sound} masking schemes (see e.g.,~\cite{RP}), $d\geq t$.
Leakage squeezing~\cite{HM:WISTP-11} aims at optimal values for $d$ (i.e., $d \gg t$) through shares encoding by bijections.

\subsection{Concrete results for masking schemes}

A useful case study for embedded systems is when the sensitive variable $Z$ (to be defined in Sec.~\ref{sec-motivation}) fits on one byte.
With one mask ($t=2$ shares),
the leakage squeezing succeeds to increase the order $d$ of the first successful attack from $2$ (the value of $t$) to $4$ with a $[16,8,5]$ code and even to $5$ with a $(16,256,6)$ code~\cite{leakage-bis}. We recall that the $(16,256,6)$ Nordstrom-Robinson code is the binary image of the $(8,4^4,6)$ $\mathbb{Z}_4$-linear code called the octacode through the Gray map~\cite{forney-NR}; this motivates further the investigation of $\mathbb{Z}_4$-linear codes.
With two masks ($t=3$ shares),
it is possible to increase the order $d$ from $3$ (the value of $t$) to $7$ with a $[24,8,8]$ code.

The gain in terms of security is directly related to the order $d$ of the attack:
the number of observations to successfully recover the key will increase
in proportions of $\sigma^{2 d}$~\cite{DBLP:conf/ches/WaddleW04}, where
$\sigma^2$ is the variance of the measurement noise.
The overhead, in terms of cost, is null or negligible (especially in hardware implementations of masking):
it merely consists of calls to the bijections $F_i$ ($0 \leq i < t$). In implementations that make use of look-up tables (e.g., block ciphers), the bijections can be merged with the memory that is looked-up, resulting in no additional cost. For other applications, the cost is that of evaluating the bijections $F_i$.

In some cases, linear $F_i$ can be preferred.
Indeed, they provide simultaneously a protection against leakages
\begin{itemize}
\item in \emph{value},
when the leak comes from the values stored in registers, in the so-called Hamming weight model, and
\item in \emph{distance},
when the leak comes from the differences between the values stored in registers and the values stored previously in the same registers, in the so-called Hamming distance model~\cite{HM:JCEN14}. 
\end{itemize}
Moreover, in general, they might be easier to implement than non-linear ones.
Especially, in the context of high-order masking~\cite{fse2012_carlet}, the linear operations are often ignored (since faster than field multiplications) when computing a time complexity. Eventually, linear bijections allow to relate the $t$ bijections $F_i$ ($0 \le i < t$) to a $t$-CIS code (See Section~\ref{sub-tCI}).

\subsection{Outline of the paper}

The material is organized as follows. Section~\ref{sec-motivation} describes and discusses the security motivation.
Section~\ref{sec-notation} collects the necessary notations and definitions.
Section~\ref{sec-boolean_functions} develops the interplay with (vectorial) Boolean functions.
Section~\ref{sec-Z4} studies $\Z_4$-codes.
Section~\ref{sec-asymptotics} shows that arbitrarily long $t$-CIS codes exist for given $t$.
Section~\ref{sec-tCISalgo} describes and runs the CISness testing algorithm.
Section~\ref{sec-numerical_examples} gives numerical examples of $t$-CIS codes for $t=3,4, \dots, 256$.
Section~\ref{sec-cstr} gives classical construction methods based on respectively, quasi-cyclic codes and the building up construction of \cite{Kim01, 2CIS} as well as
a formula similar to a mass formula.
Section~\ref{sec:class-CIS} gives a classification of $3$-CIS codes of length up to $12$.
Finally, Section~\ref{sec-concusion} concludes the paper with some open problems.

\section{Motivation}
\label{sec-motivation}

\subsection{Boolean masking with leakage squeezing}

Any embedded system leaks information about the data it processes and should thus be protected against side-channel attacks that are able to exploit such a leakage.
The attack targets are ``sensitive variables'', i.e., varying data, that depend on a secret key concealed within the device.
Usually, for implementation reasons, the sensitive variable $Z$ has a size of $k$ bits, where $k$ is a typical word length of computing machines, i.e.,
$4$ (\emph{nibble}),
$8$ (\emph{byte}),
$16$ (\emph{word}) or
$32$ (\emph{double word}) bits, although some custom circuits can use different sizes better suited to the algorithm.
The variable $Z$ depends on $k$ bits known by the attacker (e.g., from the input or the output of the algorithm) and of $k$ bits of a secret key.
A usual situation is when $Z$ is the exclusive-or of a public data and a part of secret key;
like in the AddRoundKey step of the first round of the Advanced Encryption Standard (AES~\cite{website-fips197}) block cipher.
The attack consists in guessing $Z$ for all $2^k$ possible choices of the $k$-bit key part
(in the rest of the article, this \emph{key part} of $k$ bits is simply called a \emph{key}),
and selecting the key that maximizes the mutual information between the observed leakage and the value of $Z$.

To prevent such attacks, countermeasures are applied.
For instance, high-order masking consists in splitting $Z$ into $t>1$ shares $S_i$ ($0 \leq i < t)$, in such a way:
\begin{itemize}
\item $Z$ is a deterministic function of all the $S_i$, but
\item $Z \independent (S_i)_{i \in I}$ if $|I| < t$ (where $\independent$ means ``statistically independent'').
\end{itemize}
A high-order masking scheme that satisfies these properties for every possible sensitive variable $Z$ is qualified \emph{sound}.

A convenient and widely studied high-order masking is the additive Boolean Masking~\cite{DBLP:conf/ctrsa/SchrammP06},
where the sharing is done in the group $(\mathbb{F}_2^k, \oplus)$;
the random shares $S_i$ are drawn uniformly in $\mathbb{F}_2^k$ while satisfying the constraint $Z = \bigoplus_{i=0}^{t-1} S_i$.
Notice that the high-order masking scheme of Schramm and Paar~\cite{DBLP:conf/ctrsa/SchrammP06} is sound only at order $2$~\cite{DBLP:conf/ches/CoronPR07}.
Recently, Rivain and Prouff~\cite{RP} (2010) and Coron~\cite{htable} (2014) have given constructions for sound high-order masking schemes at any order.

Such high-order masking~\cite{RP} leaks no information about $Z$,
hence protects unconditionally the secret keys,
when strictly less than $t$ shares are exploited simultaneously.
Recently, it has been warned that this condition holds only if the scheme is implemented properly, e.g., without unintended interactions between the shares.
For instance, conditional glitches (spurious transitions occurring in hardware circuits) can constitute unintended interactions~\cite{DBLP:conf/ches/MangardS06}.
But in reaction, masking schemes have been upgraded~\cite{DBLP:conf/ches/ProuffR11} to face this risk also.

The leakage squeezing~\cite{HM:WISTP-11} aims at reducing as much as possible the leakage when the attacker is able to gather the $t$ shares.
It introduces an encoding of the shares that allows to decorrelate them as much as possible.
The goal is to make high-order attacks difficult.
Notice that leakage squeezing can apply to any high-order masking schemes;
for the sake of clarity, we focus in this paper on leakage squeezing on additive Boolean masking.

Their formalization requires the introduction of the notion of leakage function $L_i$ $(0 \leq i < t)$.
In the optimal case for the attacker, each share $S_i$ is leaked independently through $L_i: \mathbb{F}_2^k \to \mathbb{R}$.
The leakage function $L_i$ can be written as the composition:

$$L_i(S_i) =
\fbox{\small $\begin{array}{c} \text{attacker's} \\ \text{function} \end{array}$} \, \circ
\fbox{\small $\begin{array}{c} \text{device's}   \\ \text{function} \end{array}$} \, \circ
\fbox{\small $\begin{array}{c} \text{defender's} \\ \text{function} \end{array}$} \, (S_i) \enspace.$$

Typical examples are:
\begin{itemize}
\item Defender's function: a function $F_i: \mathbb{F}_2^k \to \mathbb{F}_2^k$, which must be bijective, since at the end of the computation one must recover the shares $S_i$ from their image by $F_i$;
\item Device's   function: it is mapping $\ell_i$ from $\F_2^k$ to $\R$ that represents the \emph{transduction} from ``bits'' to the ``side-channel physical quantity'' (e.g., Volts if a voltage is measured, or Amperes if a current is measured, etc.).
From a mathematical perspective, it is a pseudo-Boolean function of unit numerical degree,
i.e., an affine function of the input bits.
If efforts are done to balance the hardware, then $\ell_i$ can be the Hamming weight.
\item Attacker's function: an application $H_i: \mathbb{R} \to \mathbb{R}$ that raises the leakage's degree, typically the ``power'' function $x \mapsto x^{p_i}$, for some $p_i \geq 1$.
\end{itemize}

\medskip

The attacker exploits this leakage by computing the \textbf{optimal combination}, i.e., the product $\prod_{i=0}^{t-1} L_i(S_i)$~\cite{DBLP:journals/tc/ProuffRB09}.
This combination by product is optimal in the sense that it minimizes the effect of the noise.
The optimality of this combination function has been conjectured in~\cite{DBLP:journals/tc/ProuffRB09}, but we prove it for the first time in this article. Recently, the same result has been proved independently  by a team from Northeastern University (Boston, USA) using a different path~\cite{cryptoeprit:2014:433}.

Typically the $L_i(S_i)$ are measured as a random variable $L_i(S_i)+N_i$
affected by a centered noise $N_i$ of variance $\sigma^2_i$.
For instance, when $p_i=1$, it is a customary hypothesis to take $N_i \sim \mathcal{N}(0,\sigma^2_i)$, i.e., an additive white Gaussian noise (AWGN).
Any combination of leakage functions can be written, in a general way, as a polynomial in $\mathbb{R}[L_0(S_0), \cdots, L_{t-1}(S_{t-1})]$:
$\sum_{\vec{\alpha}=(\alpha_i) \in \mathbb{N}^t} \beta_{\vec{\alpha}} \prod_{i=0}^{t-1} L_i(S_i)^{\alpha_i}$,
where the $\beta_{\vec{\alpha}}$ are real coefficients (possibly equal to zero).
As the attacker does not know the shares $S_i$ but only their leakages $L_i(S_i)$,
(s)he simply checks whether there is a dependence, in average, with $Z=z$.
Now,
\begin{align*}
& \mathbb{E}\left\lbrack
\textstyle
\sum_{\vec{\alpha}=(\alpha_i) \in \mathbb{N}^{t}} \beta_{\vec{\alpha}} \prod_{i=0}^{t-1} L_i(S_i)^{\alpha_i}
\mid Z=z \right\rbrack \\
&= \textstyle
\sum_{\vec{\alpha}=(\alpha_i) \in \mathbb{N}^{t}} \beta_{\vec{\alpha}} \mathbb{E}\left\lbrack \prod_{i=0}^{t-1} L_i(S_i)^{\alpha_i} \mid Z=z \right\rbrack \\
&~~~~\text{(by the linearity of the expectation)} \\
&= \textstyle
\sum_{\vec{\alpha}=(\alpha_i) \in \mathbb{(N^\star)}^{t}} \beta_{\vec{\alpha}} \mathbb{E}\left\lbrack \prod_{i=0}^{t-1} L_i(S_i)^{\alpha_i} \mid Z=z \right\rbrack \enspace \\
&~~~~\text{(by the soundness property of the masking)}
\end{align*}
This means all the exponents $\alpha_i$ in the combination function must be strictly positive.
The smallest effect of the noise occurs when they are all equal to $1$,
and the terms of higher degree simply add more noise, thus shall not be taken into consideration.
Consequently, the optimal combination results from the choice $\forall \vec{\alpha} \neq (1,1,\cdots,1), \beta_{\vec{\alpha}}=0$.
By convention, we set $\beta_{(1,1,\cdots,1)}=1$.

The dependence between the optimal combination of the leakage of the $t$ shares and the sensitive variable $Z$ writes:
\[
\begin{array}{l}
\mathbb{E}\left\lbrack \prod_{i=0}^{t-1} L_i(S_i) \mid Z=z \right\rbrack  \\
=\frac1{2^{k(t-1)}} \sum_{s_1, \cdots, s_{t-1}} L_0 ( \underbrace{z \oplus \bigoplus_{i=1}^{t-1} s_i}_{=s_0} ) \cdot \prod_{i=1}^{t-1} L_i(s_i) \\
=2^{-k(t-1)} \bigotimes_{i=0}^{t-1} L_i (z), 
\label{eqn-E}
\end{array}
\]
where $\otimes$ represents convolution.
The attacker can successfully build a distinguisher between the (only) correct and the (many) incorrect key hypotheses if and only if this quantity depends on $z$.
As a corollary, the attack fails if and only if $\bigotimes_{i=0}^{t-1} L_i (z)$ is a constant.
Recall that $L_i$ is typically written as $L_i = H_i \circ \ell_i \circ F_i$;
If $F_0$ is linear, so is $F_0^{-1}$, and $\bigotimes_{i=0}^{t-1} L_i(z) = \bigotimes_{i=0}^{t-1} L'_i(z')$, where $L'_i=L_i \circ F_0^{-1}$ and $z'=F_0(z)$.
So $\bigotimes_{i=0}^{t-1} L_i$ is a constant if and only if $\bigotimes_{i=0}^{t-1} L'_i$ is.
This means that $F_0$ can be chosen equal to the identity $\mathsf{Id}$, which is customarily assumed in articles about leakage squeezing~\cite{HM:WISTP-11,leakage,leakage-bis,3CIS,HM:JCEN14}.
We will also make this assumption in the sequel.

Without leakage squeezing, i.e., for all $i \in \{0, \cdots, t-1\}, F_i = \mathsf{Id}$,
the convolution product $\bigotimes_{i=0}^{t-1} L_i(z)$ depends on $z$ whatever the values $p_i$ are,
and in particular also if they have their smallest possible values $p_i=1$.
Now, as shown in~\cite{leakage}, the attack becomes all the more difficult as $\sum_i \text{d}^\circ(H_i) = \sum_i p_i$, where $\text{d}^\circ(H_i)$ denotes the degree of $H_i$, is high.
Logically, this quantity is called the order of the attack%
\footnote{To be more rigorous, we shall consider for the order of the attack the sum of the algebraic degrees of the pseudo-Boolean functions $H_i \circ \ell_i$.
However, it happens to coincide with $\sum_i \text{d}^\circ(H_i)$ because the $\ell_i$ have unit numerical degree.}.
Raising this quantity from $t$ to greater values $d$, is the topic
of~\cite{leakage-bis} for $t=2$ and
of~\cite{3CIS}        for $t=3$.
In Section 3.3 of~\cite{3CIS}, it is explained that non-trivial bijections $F_i$ manage to increase the order of the attack from $t$ to $d$.
The conditions on the $F_i$ are expressed in Sec.~\ref{sec-boolean_functions}.
An example (using in advance the results from Sec.~\ref{sec-boolean_functions} and~\ref{sec-numerical_examples}) is completely explicited below.

In this article, we tackle the general case (arbitrary $k$ and $t$), i.e., finding the $t$ bijections $F_i: \F_2^k\to\F_2^k$ that maximize $d$ for a given pair $(k,t)$.
This comes down to achieving the highest security level for a given overhead.
An equivalent problem would be to fix $k$ and $d$, and to find the smallest $t$.
This comes down to minimizing the overhead for a given security level.

\subsection{Example for $k=8$ and $t=3$}
\label{sub-k8_t3}

\begin{figure}
\begin{center}
\includegraphics[width=0.9\linewidth]{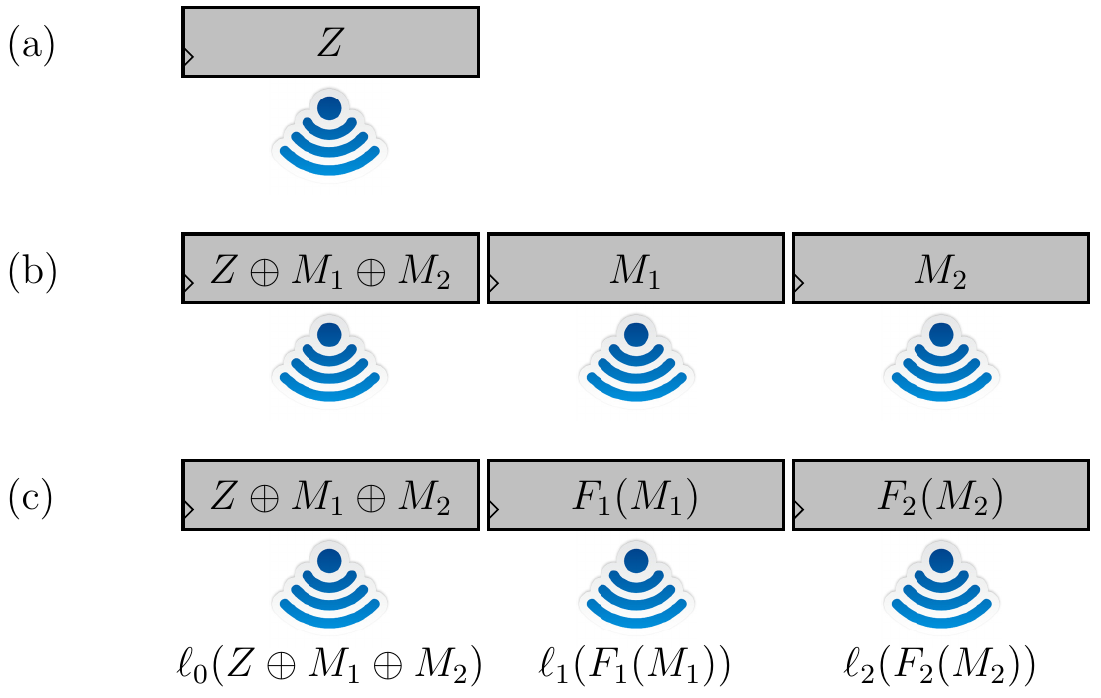}
\end{center}
\caption{Illustration of unintentional side-channel leakage,
(a) without protection,
(b) with a $3$-share additive Boolean masking,
(c) with leakage squeezing of order $2$, using two bijections $F_i:\F_2^k\to\F_2^k$ ($i\in\{1,2=t-1\}$).}
\label{fig-LS3}
\end{figure}

An algorithm such as AES manipulates sensitive variables $Z\in\F_2^k$ that are bytes ($k=8$).
Without care, $Z$ leaks through various side-channels (e.g., the electromagnetic field) emitted by the register it resides in.
The register is the hardware resource that memorizes the state of a circuit, e.g., of the current state of an iterative block cipher such as the AES.
This is sketched in Fig.~\ref{fig-LS3}(a):
the register is the grey box with the tiny triangle on its left side.
It means that at every clock cycle, this register is sampling a new sensitive variable $Z$, and consequently leaks some information about $Z$ (which is symbolized by blue waves).
The additive Boolean countermeasure consists in splitting $Z$ into several shares (here $t=3$)
using random numbers (here $M_1, M_2\in\F_2^k$).
Thus, the attacker must collect and combine $3$ leakages, as shown in Fig.~\ref{fig-LS3}(b).
We have nonetheless that $(\ell_0 \otimes \ell_1 \otimes \ell_2)(z)$, quantity proportional to Equation~\eqref{eqn-E}, depends on $z$;
for instance, when all $\ell_i$ ($0 \leq i <3$) are the Hamming weight function%
\footnote{%
It is shown in the Equation~(6) of~\cite{leakage}, located in the proof of Theorem 2,
that $\forall z\in\F_2^k, \forall t>0$,
$\left( \bigotimes_{i=0}^{t-1} w_H \right)(z) = \left(-\frac{1}{2}\right)^{t-1} \left( w_H(z) + \frac{k}2 \left( (-k)^{t-1}-1\right)\right)$.}
then $2^{-2k} (w_H \otimes w_H \otimes w_H)(z)=w_H(z)/4 + (k-1) k (k+1) / 8$.
%

Now, the leakage squeezing (with linear bijections) consists in finding two linear permutations of $\F_2^k$,
denoted by $F_1$ and $F_2$, such that:
\begin{itemize}
\item not only $\left( \ell_0 \otimes (\ell_1 \circ F_1) \otimes (\ell_2 \circ F_2) \right)(z)$ does \underline{not} depend on $z\in\F_2^k$,
\item but also such that $\left( \ell_0^{p_0} \otimes (\ell_1 \circ F_1)^{p_1} \otimes (\ell_2 \circ F_2)^{p_2} \right)(z)$ does \underline{not} depend on $z\in\F_2^k$ for all $p_i \geq 1$ ($i\in\{0,1,2\}$) such as $p_1+p_2+p_3 \leq d$, for $d$ as large as possible.
\end{itemize}
The bijections are applied to the shares $M_1$ and $M_2$, as can be seen from Fig.~\ref{fig-LS3}(c).

As proven later in Theorem~\ref{thm-two-permut1},
the condition of $F_1$ and $F_2$ is that they form a Correlation Immune Pair,
i.e., that the code
\begin{equation*}
C(F_1,F_2)=\{ (M_1\oplus M_2,F_1(M_1),F_2(M_2)) |\, M_1,\,M_2 \in \F_2^k\}
\end{equation*}
has dual distance at least $d+1$.

Then, as mentioned in the introduction, such a code is better researched by looking at its dual.
As proven later in Theorem~\ref{thm-two-permut2},
the code $C(F_1,F_2)$ has dual distance at least $d+1$ if and only if the $[3k,k]$ linear code
\begin{equation*}
C(F_1,F_2)^{\perp}=\{ (u,G_1(u),G_2(u)) |\, u \in \F_2^k\}
\end{equation*}
is $3$-CIS and has minimum distance at least $d+1$.
In this equation,
$G_1=(F_1^*)^{-1},\,G_2=(F_2^*)^{-1}$ where $F^*$ denotes the adjoint operator of $F$, that is, the operator whose matrix is the transpose of that of $F.$
Equivalently, $F_1=(G_1^{-1})^*=(G_1^*)^{-1}$ and $F_2=(G_2^{-1})^*=(G_2^*)^{-1}$.

This is where our CISness test algorithm comes into play.
The two functions $F_1$ and $F_2$ can be found in three steps, detailed hereafter.
\begin{enumerate}
\item
The best known linear code $[tk,k]$ is checked for CISness.
We have found in Sec.~\ref{sub-3CIS} that the best known linear code $[24,8,8]$ is $3$-CIS.
So we know that masking with second-order leakage squeezing allows to resist to attacks of order $3$, $4$, $5$, $6$ and $7$.
The first attack to succeed is at order $8$.

Note that in case the best known linear code had not been CIS,
we would been obliged to fall back on non-optimal codes
(that in turn are tested for CISness)
until a CIS code of parameters $[tk,k]$ is found.
\item
The code is written under systematic form as $\{ (u,G_1(u),G_2(u)) |\, u \in \F_2^k\}$.
Its generating matrix writes as
$(I_8 ~ L_1 ~ L_2)$,
where:
\begin{equation*}
{\tiny L_1 =
\left(
\begin{array}{cccccccc}
1 & 0 & 0 & 0 & 0 & 1 & 0 & 1 \\
1 & 0 & 0 & 0 & 0 & 1 & 1 & 1 \\
1 & 0 & 1 & 1 & 1 & 0 & 0 & 1 \\
1 & 0 & 1 & 1 & 1 & 0 & 1 & 0 \\
1 & 0 & 1 & 1 & 1 & 1 & 1 & 0 \\
0 & 1 & 1 & 0 & 0 & 1 & 1 & 1 \\
0 & 1 & 0 & 1 & 0 & 1 & 1 & 1 \\
0 & 1 & 0 & 0 & 1 & 0 & 0 & 0 \\
\end{array}
\right)}%
,\end{equation*}

\begin{equation*}
{\tiny L_2 =
\left(
\begin{array}{cccccccc}
0 & 1 & 1 & 0 & 0 & 1 & 1 & 0 \\
0 & 0 & 0 & 1 & 1 & 0 & 1 & 0 \\
1 & 1 & 1 & 1 & 0 & 1 & 1 & 1 \\
1 & 0 & 1 & 0 & 0 & 0 & 0 & 1 \\
1 & 1 & 0 & 1 & 1 & 0 & 1 & 0 \\
1 & 1 & 1 & 1 & 0 & 0 & 0 & 1 \\
1 & 1 & 1 & 1 & 1 & 1 & 1 & 0 \\
1 & 0 & 0 & 0 & 1 & 1 & 1 & 1 \\
\end{array}
\right)}
\end{equation*}
are two matrices of maximal rank $k=8$.
\item
Now, $F_1$ and $F_2$ are deduced from $G_1$ and $G_2$.
Let us denote by $\transpose{L}$ the transpose of the square $k\times k$ matrix $L$.
We get that $F_1$ maps $M_1$, seen as a column, to $F_1(M_1)=(\transpose{L_1})^{-1} \cdot M_1$,
and    that $F_1$ maps $M_1$, seen as a column, to $F_2(M_2)=(\transpose{L_2})^{-1} \cdot M_2$.
The two matrices involved are given below:
\begin{equation*}
(\transpose{L_1})^{-1} =
\left(
\tiny
\begin{array}{cccccccc}
0 & 1 & 1 & 1 & 1 & 0 & 1 & 1 \\
1 & 0 & 0 & 0 & 0 & 0 & 1 & 1 \\
1 & 1 & 0 & 0 & 1 & 0 & 0 & 1 \\
0 & 1 & 1 & 1 & 1 & 1 & 0 & 1 \\
1 & 1 & 0 & 0 & 1 & 1 & 0 & 0 \\
0 & 1 & 0 & 1 & 1 & 0 & 0 & 0 \\
0 & 1 & 1 & 0 & 1 & 0 & 0 & 0 \\
0 & 1 & 1 & 1 & 0 & 0 & 0 & 0 \\
\end{array}
\right)%
,
\end{equation*}

\begin{equation*}
(\transpose{L_2})^{-1} =
\left(
\tiny
\begin{array}{cccccccc}
0 & 0 & 1 & 0 & 1 & 1 & 1 & 1 \\
0 & 1 & 1 & 1 & 1 & 1 & 1 & 1 \\
0 & 0 & 1 & 0 & 0 & 1 & 0 & 1 \\
0 & 0 & 0 & 1 & 1 & 0 & 0 & 1 \\
0 & 1 & 1 & 1 & 1 & 0 & 0 & 1 \\
1 & 1 & 0 & 0 & 1 & 0 & 1 & 1 \\
1 & 1 & 1 & 1 & 1 & 0 & 0 & 0 \\
1 & 1 & 1 & 1 & 0 & 1 & 1 & 0 \\
\end{array}
\right).
\end{equation*}

\end{enumerate}

\section{Notations and definitions}
\label{sec-notation}

\subsection{Binary codes}
For basic definitions of codes, we refer to~\cite{HufPle},~\cite{JoyKim},~\cite{MS77}.
Let $C$ be a binary linear code. Its parameters are formatted as $[n,k,d]$ denoting length, dimension, and minimum distance. By an {\bf unrestricted code} we shall mean a possibly nonlinear code.
 The dual $C^{\perp}$ of a linear code $C$ is understood to be with respect to the standard inner product.

A binary (unrestricted) code $C$ of length $n$ is called {\bf systematic} if there  exists a  subset $I$ of $\{1,\cdots,n\}$ called an
{\bf information set} of $C$, such that every possible tuple of length $|I|$ occurs in exactly one codeword within the specified coordinates $x_i;\; i\in I$. Every non trivial linear code is systematic in this sense, since it admits a generator matrix in which all the vectors of the canonical basis of $\F_2^k$ are columns. The generator matrix of a $[tk,k]$ code is said to be in {\bf systematic form} if these columns are at the first $k$ positions, that is, if it is blocked as $(\mathsf{I}_k|A)$ with $\mathsf{I}_k$ the identity matrix of order $k.$
We call a systematic code of length $tk$ which admits $t$ pairwise disjoint information sets a {\bf $t$-CIS (unrestricted) code}.

The {\bf Hamming weight} $w_H(z)$ of a binary vector $z$ is the number of its nonzero entries. The {\bf Hamming distance} $d_H(x,y)$ of two binary vectors $x,y$  is defined as the weight of their sum $d_H(x,y)=w_H(x+y).$

A binary code is said to be {\bf $s$-quasi-cyclic} if it is wholly invariant under $T^s$ where $T$ stands for the shift operator and the index $s$ divides $n.$ Such codes have a natural module structure over the ring $\F_2[x]/(x^m-1),$ where $m=n/s.$ The code is said to be $1$-generator if it has a single generator as such a module.

\subsection{Boolean functions}\label{bf}


A permutation $F$ of $\F_2^k$ is any bijective map from  $\F_2^k\rightarrow \F_2^k,$ also called a {\bf vectorial Boolean function}. Let $a, b \in \F_2^k$. The {\bf Walsh-Hadamard} transform of $F$ at $(a,b)$ is defined by
$$ W_F(a,b)=\sum_{x\in \F_2^k}(-1)^{ a\cdot x + b\cdot F(x)},$$ where $a\cdot x$ denotes the usual scalar product of vectors $a$ and $x$.

 If $f$ is a Boolean function with domain $\F_2^k$ and range $\mathbb F_2$, then
  the {\bf Fourier transform} $\hat{f}$ of $f$ at $a$ is defined by
$$ \hat{f}(a)=  \sum_{x\in \F_2^k}f(x)(-1)^{a\cdot x}=\sum_{x\in {\mbox{\small{supp}}}(f)}(-1)^{a\cdot x},$$
where ${\mbox{supp}}(f)$ is the support of function $f$.

Also if $F_1$ is any permutation of $\F_2^k$ and $b$ is in $\F_2^k$, then let $f=b \cdot F_1$ so that for $x \in \F_2^k$, $f(x)=b \cdot F_1(x)$, the usual scalar product of $b$ and $F_1(x)$ as valued in $\{0,1\}$. Then the Fourier transform of $f$ at $a$ is

$$\hat{f}(a)= \widehat{b \cdot F_1}(a)=\sum_{x\in \F_2^k|b \cdot F_1(x)=1}(-1)^{a\cdot x}.$$
Considering $b \cdot F_1(x)$ as valued in $\{0,1\}\subset {\Bbb Z}$, we have $(-1)^{b \cdot F_1(x)}=1-2(b \cdot F_1(x))$. Then, we have that the Walsh-Hadamard transform value $W_{F_1}(a,b) =\sum_{x\in \F_2^k}(-1)^{b \cdot F_1(x)+a\cdot x}$ equals $\sum_{x\in \F_2^k}(-1)^{a\cdot x}-2 (\widehat{b \cdot F_1}(a))$. If $a=0$ then $\sum_{x\in \F_2^k}(-1)^{a\cdot x}=2^k$ and otherwise $\sum_{x\in \F_2^k}(-1)^{a\cdot x}=0$. Hence for $a\neq 0$ we have that $W_{F_1}(a,b) = \sum_{x\in \F_2^k}(-1)^{b \cdot F_1(x)+a\cdot x}=-2 (\widehat{b \cdot F_1}(a))$. Therefore we see that for $a \neq 0$,
\begin{equation} \label{eq:two-hats}
W_{F_1}(a,b) =0 {\mbox{ if and only if }} \widehat{b \cdot F_1}(a)=0.
\end{equation}

\subsection{Dual distance}
If $C$ is a binary code of length $n$, let  $(B_i)_{i=0,\dots ,n}$ denote its distance distribution,
that is,
$$B_i=\frac{1}{|C|}\left|\{(x,y)\in C\times C \, |\, d_H(x,y)=i\}\right|.$$
The dual distance distribution  $(B_i^\perp)_{i=0,\dots ,n}$ is the MacWilliams transform of the distance distribution, in the sense that
$$D_C^{\perp}(X,Y)=\frac{1}{|C|}D_C(X+Y,X-Y),$$
where $$D_C(X,Y)=\sum_{i = 0}^nB_i X^{n-i}Y^i, $$ denotes the {\bf distance enumerator}
and $$D_C^\perp (X,Y)=\sum_{i = 0}^nB_i^\perp X^{n-i}Y^i.$$ The {\bf dual distance} of $C$ is
the smallest $i>0$ such that $B_i^{\perp}\neq 0.$ When $C$ is
linear, it is the minimum distance of $C^{\perp}$, since $D_C^\perp (X,Y)=D_{C^\perp}(X,Y)$.

\subsection{$\Z_4$-codes}
Recall that the Gray map $\phi$ from $\Z_4$ to $\F_2^2$ is defined by \\
$$\phi(0)=00, \ %
  \phi(1)=01, \ %
  \phi(2)=11, \ %
  \phi(3)=10.$$
This map is extended componentwisely from $\Z_4^n$ to ${\Bbb
F}_2^{2n}$;
It is referred to as the \emph{binary image}.
A {\bf $\Z_4$-linear code} of length $n$ is a $\Z_4$-submodule of
$\Z_4^n$ and is called a $\Z_4$-code for short. The {\bf binary image} $\phi(C)$ of a $\Z_4$-code $C$ is
just $\{\phi(c)|\,c\in C\}.$ In general, a $\Z_4$-code $C$ is of type
$4^k2^l$ if $C\cong \Z_4^k \Z_2^l$ as additive groups. A $\Z_4$-code is called {\bf free} if $l=0.$
The parameters of a $\Z_4$-code are hereby formatted as $(n,|C|, d_L(C)),$ where $d_L(C)$ denotes the minimum distance of the binary image of $C$, called the Lee distance.
\section{Boolean functions}
\label{sec-boolean_functions}

\subsection{Characterization of $3$-CIS codes}

For simplicity's sake we will mainly consider the case $t=3.$
\medskip

A pair $(F_1,F_2)$ of permutations of $\F_2^k$ forms a {\bf Correlation Immune Pair} (CIP) of strength $d$ if and only if
 for every $(a,b,c)$ such that $a,b,c \in \F_2^k$, $a\neq 0$, and $w_H(a)+w_H(b)+w_H(c)\leq d$, we have $\widehat{b\cdot F_1}(a)=0$ or $\widehat{c\cdot F_2}(a)=0$, equivalently
 $W_{F_1}(a,b) =0$ or $W_{F_2}(a,c) =0$ by Equation~(\ref{eq:two-hats}).

This notion has actually already been introduced in a slightly different way in~\cite{3CIS}.
It expresses the fact that the leakage squeezing with two masks (i.e., $t=3$ shares) and two permutations $F_1$ and $F_2$ allows to resist high-order attacks of order $d$.
We here give it the name of CIP of strength $d$.

The definition of a CIP of strength $d$ is equivalent to Condition (8) in~\cite{3CIS}, that we recall now:
\[
\label{CNS} 
\begin{array}{l}
\forall a\in \mathbb{F}_2^k, a\neq 0,\exists q,r \text{ such that } \\
\left\{
\begin{array}{l}
\HW{a}+q+r=d-1,\\ 
\forall b\in \mathbb{F}_2^k, \HW{b}\leq q \Longrightarrow \widehat{b
\cdot F_1}(a) = 0, \\
\forall c\in \mathbb{F}_2^k, \HW{c}\leq r \Longrightarrow \widehat{c
\cdot F_2}(a) = 0. \\ 
\end{array}
\right.
\end{array}
\]

The reason is as follows. For a given $a\neq 0$, we denote by $q_1$ the maximal number such that $\widehat{b\cdot F_1}(a)=0$ for every $b$ such that $w_H(b)\leq q_1$ and by $r_1$  the maximal number such that $\widehat{c\cdot F_2}(a)=0$ for every $c$ such that $w_H(c)\leq r_1$. Then there exists $b$ such that $w_H(b)=q_1+1$ and $\widehat{b\cdot F_1}(a)\ne 0$, and there exists $c$ such that $w_H(c)=r_1+1$ and $\widehat{c\cdot F_2}(a)\ne 0$. In Condition (8) recalled above, we necessarily have $q\leq q_1$ and $r\leq r_1$ (by the definition of $q_1$ and $r_1$).

If Condition (8)~\cite{3CIS} is satisfied, we have that $w_H(a)+(q_1+1)+r_1 =w_H(a)+q_1+(r_1+1)\geq w_H(a)+q+r+1=d$, and hence $w_H(a)+q_1+r_1\geq d-1$. This implies the condition in the definition of CIP. This is because given $(a,b,c)$ such that $a,b,c \in \F_2^k$, $a\neq 0$, and $w_H(a)+w_H(b)+w_H(c)\le d$, we have $\widehat{b\cdot F_1}(a)=0$ or $\widehat{c\cdot F_2}(a)=0$ since $w_H(a)+w_H(b)+w_H(c)\le d$ implies $w_H(b)\le q_1$ (and $\widehat{b\cdot F_1}(a)=0$ ) or $w_H(c)\le r_1$ (and $\widehat{c\cdot F_2}(a)=0$). Conversely, if Condition (8)~\cite{3CIS} is not satisfied then we have $w_H(a)+q_1+r_1<d-1$ for some $a\neq 0$; then choosing $b$ of weight $q_1+1$ such that $\widehat{b\cdot F_1}(a)\neq 0$ and $c$ of weight $r_1+1$ such that $\widehat{c\cdot F_2}(a)\neq 0$, we see that the condition in the definition of a CIP of strength $d$ is not satisfied. This completes the proof of the equivalence between a CIP of strength $d$ and Condition (8) in~\cite{3CIS}.







\medskip

We are now ready for the coding theoretic characterization of CIP.

\begin{thm}\label{thm-two-permut1} If $F_1,\,F_2$ are permutations of $\F_2^k$ then they form a CIP of strength $d$ if and only if the systematic code of length $3k$ and size $2^{2k}$

\begin{equation}
C(F_1,F_2)=\{ (x+y,F_1(x),F_2(y)) |\, x,\,y \in \F_2^k\} \label{eqn-two-permut1}
 \end{equation}
 has dual distance at least $d+1$.

\end{thm}

\begin{preuve}

Let $C=C(F_1, F_2)$. To find the dual distance of $C$, we recall
$$D_C^{\perp}(X,Y)=\frac{1}{|C|}D_C(X+Y,X-Y).$$
 By the definition of $D_C(X,Y)$, we first consider the distance enumerator of this code:

 \[
 \begin{array}{l}
 D_C(X,Y)= \frac {1}{|C|} \sum_{x,y,x',y'\in \F_2^k} [ \\
X^{3k-d_H(x+y,x'+y')-d_H(F_1(x),F_1(x'))-d_H(F_2(y),F_2(y'))} \\
  Y^{d_H(x+y,x'+y')+d_H(F_1(x),F_1(x'))+d_H(F_2(y),F_2(y'))} ].
  \end{array}
  \]

 Note that for every $x,y\in \F_2^k$, we have $d_H(x,y)=w_H(x+y)$, and that, for every $u\in \F_2^k$, we have
 \[\begin{array}{l}(X+Y)^{k-w_H(u)}(X-Y)^{w_H(u)} \\
 =\sum_{a\in \F_2^k}X^{k-w_H(a)}\prod_{i=1}^k ((-1)^{u_i}Y)^{a_i} \\
 =\sum_{a\in \F_2^k}(-1)^{a\cdot u}X^{k-w_H(a)}Y^{w_H(a)}.
 \end{array} \]
 Thus, combining it with the above description of $D_C(X,Y)$, we have
 \[
 \begin{array}{l} D_C(X+Y,X-Y)= \frac {1}{|C|} \sum_{x,y,x',y',a,b,c\in \F_2^k} [ \\
 (-1)^{a\cdot (x+x'+y+y')+b\cdot (F_1(x)+F_1(x'))+c\cdot (F_2(y)+F_2(y'))} \\
 X^{3k-w_H(a)-w_H(b)-w_H(c)} Y^{w_H(a)+w_H(b)+w_H(c)} ] = \\
\frac {1}{|C|} \sum_{a,b,c\in \F_2^k} [\\
  \left(\sum_{x\in \F_2^k}(-1)^{a\cdot x+b\cdot F_1(x)}\right)^2\left(\sum_{y\in \F_2^k}(-1)^{a\cdot y+c\cdot F_2(y)}\right)^2 \\
 X^{3k-w_H(a)-w_H(b)-w_H(c)} Y^{w_H(a)+w_H(b)+w_H(c)}]  .
\end{array}
\]

 Thus
 {\small
 \[\begin{array}{l}
D_C(X+Y,X-Y)=\frac {1}{|C|} \sum_{a,b,c\in \F_2^k} [(W_{F_1}(a,b)W_{F_2}(a,c))^2 \\
  X^{3k-w_H(a)-w_H(b)-w_H(c)} Y^{w_H(a)+w_H(b)+w_H(c)}].
 \end{array}
 \]
}

Hence, if $F_1$ and $F_2$ are permutations of $\F_2^k$ that form a CIP of strength $d$, then we have
$W_{F_1}(a,b)=0$ or $W_{F_2}(a,c)=0$ for every $(a,b,c)$ such that $a,b,c \in \F_2^k$, $a\neq 0$, and $w_H(a)+w_H(b)+w_H(c)\leq d$. If $a=0$ and $(b,c)\ne (0,0)$, say if $b\ne 0$, then since $F_1$ is a permutation, the coefficient of $X^{3k-w_H(a)-w_H(b)-w_H(c)} Y^{w_H(a)+w_H(b)+w_H(c)}$ is null as well.
 This implies that the dual distance of the code $C$ is at least $d+1$.

 Conversely if the dual distance of the code $C$ is at least $d+1$, then the last equation implies that $W_{F_1}(a,b)=0$ or $W_{F_2}(a,c)=0$ for every nonzero $(a,b,c)$ such that $a,b,c \in \F_2^k$, and $w_H(a)+w_H(b)+w_H(c)\leq d$.
  Therefore, the pair $(F_1, F_2)$ is a CIP of strength $d$.
\end{preuve}

\medskip

In the case of linear permutations we get back to the definition of a $3$-CIS linear code. The following theorem was proved in Section 3.6 of~\cite{3CIS}. For self-completeness, we give its proof.

\begin{thm}{\rm(\cite[Sec. 3.6]{3CIS})} \label{thm-two-permut2} If $F_1,\,F_2$ are linear permutations of $\F_2^k$, then they form a CIP  of strength $d$  if and only if the $[3k,k]$ linear code

$$C(F_1,F_2)^{\perp}=\{ (u,G_1(u),G_2(u)) |\, u \in \F_2^k\}$$ is $3$-CIS and has minimum distance at least $d+1.$

Here $G_1=(F_1^*)^{-1},\,G_2=(F_2^*)^{-1}$ where $F^*$ denotes the adjoint operator of $F$, that is, the operator whose matrix is the transpose of that of $F.$
\end{thm}

\begin{preuve} The code
$C(F_1,F_2)$ being the set of words $(x+y,F_1(x),F_2(y)),$ with $x, y \in \F_2^k,$ its dual $  C^\perp$ is the set of
words $(u,v,w)$ such that $$(x+y)\cdot u + F_1(x)\cdot v +F_2(y)\cdot w =
x \cdot (u+F_1^*(v)) +y\cdot (u+F_2^*(w))=0 $$ for every $x, y \in \F_2^k.$
Hence $C^\perp$ is the set of words $(u,v,w)$ such that $u=F_1^*(v),\,u=F_2^*(w)$ so that
$v=(F_1^*)^{-1}(u)=G_1(u),\,w=(F_2^*)^{-1}(v)=G_2(u)$. The result follows.
\end{preuve}

\subsection{Definition of a correlation-immune $t$-uple ($t$-CI) and link with a $(t+1)$-CIS code}
\label{sub-tCI}

More generally we make the following definition for $t>2.$

 The $t$-uple $F_1,\cdots,F_{t}$ of permutations of $\F_2^k$ form a {\bf Correlation Immune  $t$-uple} ($t$-CI) of strength $d$ if and only if
 for every $(a_0,\cdots,a_{t})$ such that $a_0 \neq 0$ and $w_H(a_0)+\cdots+w_H(a_{t})\leq d$, we have that
  $$\prod_{i=1}^{t}\widehat{a_i\cdot F_i}(a_0)=0.$$

A $2$-CI was defined at the beginning of Section~\ref{sec-boolean_functions} as a CIP.
  Theorems~\ref{thm-two-permut1} and~\ref{thm-two-permut2} can also be demonstrated in the case $t>2$. As argued in Sec.~\ref{sec-motivation}, this case is motivated, like the case $t=2$, by the leakage squeezing applied on a masking scheme that involves $t$ shares.

\subsection{Bounds}

The following bounds on $d$ for a $t$-CIS $[tk, k, d]$ code can be derived immediately:
\begin{enumerate}
\item $d \geq t$ : because of the partition in $t$ information subsets,
\item $d \leq tk-k+1 = (t-1)k+1$ : because of the Singleton bound,
but since only trivial MDS (maximum distance separable) codes exist over $\F_2$, we have
$d \leq (t-1)k$ for $tk>3$.
\end{enumerate}

Nonetheless, better bounds can be obtained:
\begin{prop} For large $t$, the largest minimum distance $d$ of a binary $[kt,k,d]$ code satisfies to
\begin{enumerate}
\item $d \geq kt \left(\frac12-C/\sqrt{t} + O\left(\frac1{t^{3/2}}\right)\right)$,
with $C=\sqrt{\log(2)/2}$.
\item $d \leq kt 2^{k-1}/(2^k-1)$.

\end{enumerate}
\end{prop}

\begin{preuve}
 The lower bound follows by the asymptotic version of the Varshamov Gilbert bound~\cite[p.~557]{MS77} combined with the expansion about $x=0$ of the functional inverse of the
 entropy function
 $$H^{-1}(1-x)=0.5-C \sqrt{x}+O(x^{1.5}).$$
 The upper bound follows by the Plotkin bound~\cite[p.~42]{MS77}.
\end{preuve}


\section{$\Z_4$-codes and non-linear binary codes}
\label{sec-Z4}

$\Z_4$-codes of length $n$ (i.e., $\Z_4$-submodules of $\Z_4^n$) can
be useful in the context of masking (Sec. II) in that their binary image
can have a better minimum distance $d$ (i.e., their formal duals can have a better dual distance) than binary linear codes of the same length and cardinality. It is also remarked in~\cite[page~133]{3CIS} that non-linear binary codes (that is, non-linearity of permutations $F_1$ and $F_2$) might still achieve better (in the study of masking).
 Define a free $\Z_4$-code of length $tk$ with $4^k$ codewords to be {\bf $t$-CIS} if its coordinate set can be partitioned into
 $t$ disjoint information sets. The following theorem justifies the study of $t$-CIS $\Z_4$-codes since the Gray images of these codes generate systematic $t$-CIS binary (usually, non-linear) codes.

\begin{thm} \label{thm:z4-CIS}
  Suppose that $C$ is a $t$-CIS $\Z_4$-code of length $tk$ with $4^k$ codewords so that
   $C=\{(u, G_1(u), \cdots, G_{t-1}(u)) | u\in\Z_4^k\}$,
 where $G_1, \cdots, G_{t-1}$ are $(t-1)$ permutations of $\Z_4^k$.
Then its binary image is a systematic $t$-CIS code of length $2tk,$
 and of cardinality $2^{2k}.$
Furthermore, if the minimum Lee distance of $C$ is $d+1,$ then the $(t-1)$ permutations
$F_i = \phi \circ (G_i^*)^{-1} \circ \phi^{-1}$ (with $1 \leq i \leq t-1$) of $\F_2^{2k}$
 form a $(t-1)$-CI of strength $d.$
 \end{thm}

\begin{preuve}
Follows from the properties of the dual distance of a $\Z_4$-linear code.
\end{preuve}

\begin{ex}
For the case $k=8$ and $t=2$, it has already been remarked in~\cite{leakage-bis,2CIS} that the Nordstrom-Robinson code has a better minimal distance (namely $6$) than the best known linear code (of parameters $[16,8,5]$).
The derivation of the optimal bijection $F$ (referred to as $F_1$ in this section) is obtained in~\cite{leakage-bis} by a manual partitioning of the codewords coordinates.
Now, Theorem~\ref{thm:z4-CIS} gives a method to recover the same result \emph{trivially} by knowing that the Nordstrom-Robinson code is the binary image of the octacode~\cite{forney-NR}.
The generating matrix for this code can be written under \emph{systematic} form in $\Z_4$, as
\begin{align*}
(I_4 ~ M) =
\left(
\begin{array}{cccc@{\hspace{15pt}}cccc}
1&0&0&0 & 3&1&2&1\\
0&1&0&0 & 1&2&3&1\\
0&0&1&0 & 3&3&3&2\\
0&0&0&1 & 2&3&1&1\\
\end{array}
\right)
\end{align*}
and thus the researched bijection for leakage squeezing is $F(x) = \phi\left(\left( \phi^{-1}(x)\right) {\left(\transpose{M}\right)}^{-1} \right)$.
In this equation, $x\in\F_2^8$, $\phi^{-1}(x)$ is a vector (seen as a $1\times 4$ matrix) of $\Z_4^4$, and $\phi: \Z_4^4\to\F_2^8$ is the componentwise Gray map.
\end{ex}

\begin{rem}
It is stated as an \emph{open problem} in~\cite{3CIS} whether for the cases $k\in\{4,8\}$ and $t=3$ there are better solutions than linear bijections $F_1$ and $F_2$.
Theorem~\ref{thm:z4-CIS} allows to show that $\Z_4$-linear codes are not better.
Indeed, the best $\Z_4$-linear code of parameters $(6,4^2,6)$ has minimal distance $6$; thus, by Theorem~\ref{thm:z4-CIS}, it not better than the best binary linear code of parameters $[24,8,8]$ (presented in great details in Sec.~\ref{sub-k8_t3}).
Similarly, the best $\Z_4$-linear code of parameters $(12,4^4,8)$ has minimal distance $8$; thus, by Theorem~\ref{thm:z4-CIS}, it not better than the best binary linear code of parameters $[24,8,8]$.
\end{rem}

\begin{ex}
In a recent computer search, a $\Z_4$-code with the parameters $(24,4^6,18)$ was found \cite{Kiermaier-Zwanzer-ringtables}.
Its binary image has the parameters $(48,2^{12},18)$, while the best known linear $[48,12]$ code has only the minimum distance $17.$
The $\Z_4$-code admits a partition of the positions into $4$ disjoint information sets.
In the following generator matrix, this partition is given by consecutive blocks of $6$ positions.
\[
\small{
\left(
\begin{array}{c@{\hspace{10pt}}cccccc@{\hspace{10pt}}cccccc@{\hspace{10pt}}cccccc}
100000 &023213 &301011 &132301\\
010000 &231330 &013120 &303121\\
001000 &231123 &003312 &001012\\
000100 &321233 &222323 &132032\\
000010 &322333 &330001 &321033\\
000001 &321301 &313202 &122120
\end{array}
\right)
}
\]
\end{ex}
It can be checked by using the software of \cite{F} that this code is $8$-quasi-cyclic over $\Z_4$ and that its permutation group is of order $3.$ Being $8$-quasi-cyclic over $\Z_4$ in length $24$ it can therefore be constructed by the cubic construction of \cite{LS2}, that is, be decomposed into a pair $(C_1, C_2)$, where $C_1$ is a code over $\Z_4$ of length $8$ and $C_2$ is a code of length $8$ over the Galois ring $GR(4,2)$.

\begin{ex}
We denote the Kerdock $\Z_4$-code \cite{HKCSS} of length $2^k$, whose image by the Gray map is the ($\Z_4$-linear) Kerdock code of length $2^{k+1}$, by $\mathcal{K}_{k+1}$ ($k$ odd).
Its parameters are $(2^k,4^{k+1},2^k - 2^{(k-1)/2})$.
Shortening and then puncturing the code $\mathcal{K}_{5+1}$ yields a $\Z_4$-code with the parameters $(30,4^5,26)$, while the best known linear $[60,10]$ code has only the minimum distance $25$.
Since the automorphism group acts doubly transitive on the positions of $\mathcal{K}_{k+1}$, this derivation from $\mathcal{K}_{5+1}$ is unique up to isomorphism.
We checked computationally that it admits a partition of the position into $6$ disjoint information sets. Hence $\mathcal{K}_{5+1}$ is a $6$-CIS $\mathbb Z_4$-code.
\end{ex}

\begin{ex}
The $(128,4^8,120)$ $\Z_4$-code $\mathcal{K}_{7+1}$ admits a partition of the positions into $16$ disjoint information sets. Hence $\mathcal{K}_{7+1}$ is a $16$-CIS $\mathbb Z_4$-code.
The best known linear $[256,16]$ linear code is a $[256,16,113]$ code.
\end{ex}

\begin{ex}
Shortening and then puncturing the code $\mathcal{K}_{7+1}$ yields a $\Z_4$-code $\mathcal{K}_{7+1, a}$  with the parameters $(126,4^7,118)$.
This derivation is unique up to isomorphism.
It admits a partition of the positions into $18$ disjoint information sets. Hence $\mathcal{K}_{7+1,a}$ is a $18$-CIS $\mathbb Z_4$-code.
The best known linear $[252,14]$ code has only the minimum distance $113$.
\end{ex}

\begin{ex}
Shortening $\mathcal{K}_{7+1}$ at any two positions, we get a $\Z_4$-code $\mathcal{K}_{7+1,b}$ with the parameters $(126,4^6,120)$.
Again, this derivation is unique up to isomorphism.
It admits a partition of the positions into $21$ disjoint information sets. Hence $\mathcal{K}_{7+1,b}$ is a $21$-CIS $\mathbb Z_4$-code.
The best known linear $[252,12]$-code has only the minimum distance $118$.
\end{ex}

\section{Asymptotics}
\label{sec-asymptotics}

Denote by $H(x)=-x\log_2 x -(1-x)\log_2 (1-x)$ the binary entropy function \cite[p.308]{MS77}.
In this section we show that there are long $3$-CIS codes satisfying the Gilbert-Varshamov bound for rate $1/3$ codes, that is with relative distance at least $H^{-1}(1/3).$
We begin with a well-known fact \cite[p.399]{MS77}.
\begin{lem}\label{GL}
The number of invertible $k \times k$ matrices is $\sim c2^{k^2},$ with $c\approx 0.29.$
 \end{lem}

Denote by $B(k,d)$ the number of pairs of permutations $F_1,F_2$ such that $d$ columns or less of the generator matrix of $C(F_1,F_2)$ (notation of Equation~\eqref{eqn-two-permut1} of the preceding section) are linearly dependent. A crude upper bound on this function can be derived as follows.
\begin{lem}\label{bound}

The quantity $B(k,d)$ is $\le M(k,d)$ where
 $$ M(k,d)=\sum_{j=2}^d \sum_{1\le r+s \le j} {k \choose j-r-s}{k \choose r } {k \choose s }(r+s) 2^{k(2k-2)} .$$
 \end{lem}
\begin{preuve}
The set of columns of the said matrix is naturally partitioned in three parts of size $k$ each, the three information sets of the $3$-CIS property.
Let $j$ be the size of the linearly dependent family of column vectors of  the said matrix with $j-r-s$ columns in part I, $r$ in part II, $s$ in part III. Choose two columns amongst $r+s$ to be obtained as
the sum
of $j-1$ others. Neglecting the invertibility properties we have $2k-2$ columns to choose freely in parts II and III.
\end{preuve}

\begin{lem} \label{asymp}
The quantity $M(k,d)$ is dominated by $2^{2k^2-2k} 2^{3k H(\delta)}$ when $d \sim 3\delta k$ with $0 \le \delta  \le 1/2.$
\end{lem}

\begin{preuve}
 We evaluate the inner sum in $M(k,d)$ by the Chu-Vandermonde identity
$${3k\choose j}= \sum_{0\le r,s,\le j} {k \choose j-r-s}{k \choose r } {k \choose s } .$$
Then, the outer sum $$\sum_{j=0}^d {3k\choose j}$$ is evaluated by
standard entropic estimates for binomials \cite[p.310]{MS77}. Note that $r+s\le 2k,$ a sub-exponential quantity.
\end{preuve}

We are now in position to derive the main result of this section.

\begin{prop}
 For each $\delta$ such that $H(\delta) <1/3$ there are long $3$-CIS codes of relative distance $\delta.$
\end{prop}

\begin{preuve}
 Combine Lemmas \ref{GL}, \ref{bound}, \ref{asymp} to ensure that, asymptotically,
$ |GL(k,2)|^2 \gg B(k,d)$ showing the existence of a $3$-CIS code of distance $>d,$ for $k$ large enough.
\end{preuve}

\section{$t$-CIS Partition Algorithm}
\label{sec-tCISalgo}

For an introduction to matroid theory, we refer the reader to~\cite{Edm},~\cite{O},~\cite{W}.
The notion of a matroid describes an independence system based upon sets. A {\bf matroid} is a pair $(M,I)$ such that $M$ is a finite set and $I$ is a collection of subsets of $M$ (called {\bf independent sets}) where $I$ is nonempty, any subset of a set in $I$ is also in $I$, and all maximum independent subsets contained in $A \subseteq M$ have the same size. This maximum independent set size within $A \subseteq M$ is the {\bf rank} of $A$. Additionally, the {\bf span} of $A$ is the set ${\mbox{span}}(A) = \{c\in M | {\mbox{rank}}(A) = {\mbox{rank}}(\{c\} \cup A)\}$. This definition of span is equivalent to the definition of closure in~\cite{W}.

In 1964, Edmonds gave the following result regarding the partition of a matroid~\cite{Edm}.

\begin{thm}\label{EdmThm} The elements of a matroid $M$ can be partitioned into as few as $t$ independent sets if and only if there is no subset $S$ of elements of $M$ such that $|S| > t \cdot \rank(S)$.
\end{thm}

\begin{ex} If $M$ is the matroid on the columns of a matrix $A$ over a field $F,$ induced by linear dependence, then $\rank(S)$ is simply the usual $F$-rank of $S$  in the linear algebra sense. In the present application, we take $F=\F_2$ and $A$ the $k$ by $n$ generator matrix of the code tested for CISness.
\end{ex}

\begin{prop}
\label{prop-exponential}
State-of-the-art algorithms (such as \cite[Appendix~A]{3CIS} and~\cite[Section VII]{frei}) that test
for all potential partitions into $k$ information sets of the coordinate set are of \emph{exponential} complexity.
\end{prop}
\begin{preuve}
The state-of-the-art algorithms to test CISness are of complexity $$\prod_{\tau=2}^t {\tau k \choose k} =
\frac{(tk)!}{k!^t}=\frac{n!}{(\frac{n}{t}!)^t}, {\mbox{ where }} n=tk \enspace.$$
Using Stirling's approximation $\ln(n!) = n \ln n - n + O(\ln n)$,
we find that the logarithm of the complexity is $n \ln n - n - n \ln(n/t) + n + O(\ln n) =
n \ln t + O(\ln n)$.
Thus an exponential complexity in $t^n$.
\end{preuve}

The paper~\cite{Edm} sketches a polynomial time algorithm for obtaining the partition described in Theorem~\ref{EdmThm}.
A more precise execution time estimate is $O(n^3)$ \cite{K73}. Thus this algorithm based on matroid theory has an improved execution time compared state-of-the-art algorithms (see Prop.~\ref{prop-exponential}).

 Adapted from the theory in~\cite{Edm}, we obtain the following algorithm which given any linear $[tk,k]$ code determines whether it is $t$-CIS.  If it is $t$-CIS, then a partition of the columns is output.  If it is not, then a set $S$ of the columns of the generator matrix violating Theorem~\ref{EdmThm} is output.  In the algorithm we will routinely use the following terms.  Given a $[tk,k]$ code $C$ with generator matrix $G$, the set $M$ will denote the column indices (the integers from $1$ to $tk$).  We will say the subset $I$ of $M$ is independent (resp. dependent) if the corresponding columns are independent (resp. dependent).  Similarly, we will denote the $\rank(I)$ and ${\mbox{span}}(I)$ with respect to the corresponding columns.  In particular, ${\mbox{span}}(I)$ denotes the subset of indices in $M$ that are spanned by the subset $I$.  \\

\medskip

\noindent
{\bf $t$-CIS Partition Algorithm:}  An algorithm to determine if a given linear code is $t$-CIS.\\

\noindent
Input:
Begin with a binary $[tk,k]$ code $C$.\\

\noindent
Output: An answer of ``Yes'' if $C$ is $t$-CIS (along with a column partition) and an answer of ``No'' if not (along with a set of columns violating Edmonds' Theorem).

\medskip

\begin{enumerate}
\item Let $\{I_1, \dots, I_t \}$ be a set of labeled disjoint independent subsets of $M$.  (Note that each $I_i$ ($1 \le i \le t$) can be randomly assigned to each have order $1$, or one may be given the first $k$ indices of a standard form matrix $G$.)
\item Select $x \in M \setminus  \bigcup_{1\leq i \leq t}I_i$.
\item While $\bigcup_{1\leq i \leq t}I_i \subsetneq M$ do:
\begin{enumerate}
\item Initialize $S_0 :=M$.  For $j >0$, recursively define $S_j := {\mbox{span}}(I_{j'} \cap S_{j-1})$, where $j'=((j-1)\text{ mod }t)+1$.  The modulus is necessary for the indices of the independent sets. Initialize $j:=0$.
\item For the current value of $j$ check that $|S_j| \leq t \cdot \rank(S_j)$.  If the inequality is false (it is immediately clear that Theorem~\ref{EdmThm} is violated), then exit the while loop and output the set $S_j$ with an answer of ``No.''
\item If $x \in S_j$, then set $j:=j+1$ and go back to b).
\item If $x \notin S_j$, then check if $I_{j'} \cup \{x\}$ is independent.  If so then replace $I_{j'}$ with the larger independent set and repeat the while loop with a new $x \in M \setminus  \bigcup_{1\leq i \leq t}I_i$.
\item If $I_{j'} \cup \{x\}$ is dependent, then find the unique minimal dependent set $C \subset I_{j'} \cup \{x\}$ (accomplished by solving the matrix equation associated with finding the linear combination of columns in $I_{j'}$ that sum to $x$).
\item Select any $x' \in C \setminus S_{j-1}$ and replace $I_{j'}$ with $I_{j'}\cup \{x\} \setminus \{x'\}$, then set $x:=x'$ and repeat the while loop.
\end{enumerate}
\item End while loop.  If the while loop was not exited early, then output the partition $\{I_1, \dots, I_t \}$ of $M$ and answer ``Yes.''
\end{enumerate}

\section{Numerical Examples}
\label{sec-numerical_examples}

The command $BKLC(GF(2),n,k)$ from the computer package Magma \cite{magma} means the best known binary linear $[n,k]$ code as per \cite{Grassl}.
The table captions are as follows.
\begin{itemize}
\item bk= obtained by the command $BKLC(GF(2),n,k)$ from Magma.
\item bk*= same as bk with successive zero columns of the generator matrix replaced in order by successive columns of the identity matrix of order $k.$
Trivially the generator matrix of bk has $<k$ zero columns.
\item qc= quasi-cyclic.

\end{itemize}

\subsection{$3$-CIS codes}
\label{sub-3CIS}
The following tables show that all $3$-CIS codes of dimension $3$ to $85$ have the best known minimum distance among all linear $[n,k]$ codes, and in fact the best possible minimum distance for $n\le 36.$ We have checked that the best known linear $[132, 44, 32]$ code in the Magma database~\cite{magma} is not 3-CIS.

{\tiny
\[
\begin{tabular}{|l|l|l|l|l|l|l|l|l|l|l|l|l|l|l|}
 \hline
$n  $& 6 &  9 & 12& 15&  18& 21& 24& 27& 30& 33 & 36 & 39\\ \hline
$k  $&2 & 3&  4  & 5& 6&  7 & 8 & 9&10&11& 12& 13 \\ \hline
$d$&   4 & 4 & 6&7&8&8&8&10&11 &12 &12 &12  \\ \hline
code &    qc& qc& bk & bk & bk & bk* &bk*&bk& bk& bk& bk*& bk* \\
\hline
\end{tabular}
\]
}
{\tiny
\[
\begin{tabular}{|l|l|l|l|l|l|l|l|l|l|l|l|l|l|l|}
 \hline
$n  $& 42&  45 & 48& 51&  54& 57& 60& 63& 66& 69 & 72 & 75\\ \hline
 $k  $& 14& 15&  16 & 17& 18&  19 & 20& 21& 22& 23& 24& 25 \\ \hline
$d$&   13& 14 & 15& 16& 16& 16& 17& 17&18 & 18 &20 & 20\\ \hline
code &    bk& bk& bk & bk & bk* & bk* &bk&bk& bk& bk& bk*&bk \\
\hline
\end{tabular}
\]
}

{\tiny
\[
\setlength{\tabcolsep}{1 mm}
\begin{tabular}{|l|l|l|l|l|l|l|l|l|l|l|l|l|l|l|l|l|l|l|l|l|l|l|}
 \hline
$n  $& 78& 81 & 84& 87& 90 & 93& 96 & 99& 102& 105 & 108 &111&114&117&120   \\ \hline
 $k  $& 26& 27&  28 & 29& 30& 31 & 32& 33 & 34 & 35& 36&37& 38&39&40        \\ \hline
$d$& 20  &22  &22 &24 &24 &24 &24 &24 & 24& 26 &26 &26 &27 & 28&28          \\ \hline
code &    bk*& bk& bk* & bk & bk* & bk* &bk&bk& bk& bk& bk*&bk& bk& bk & bk \\ \hline
\end{tabular}
\]
}

{\tiny
\[
\setlength{\tabcolsep}{1 mm}
\begin{tabular}{|l|l|l|l|l|l|l|l|l|l|l|l|l|l|l|l|l|l|l|l|l|l|l|}
\hline
$n$ &123&126&129&132&135&138&141&144&147&150&153&156&159&162 \\ \hline
$k$ & 41& 42& 43& 44& 45& 46& 47& 48& 49& 50& 51& 52& 53& 54 \\ \hline
$d$ & 29& 31& 32&  ?& 32& 32& 32& 32& 34& 34& 33& 34& 34& 35 \\ \hline
code&bk*&bk*&bk*&  ?&bk*&bk*&bk*&bk*& bk&bk*& bk&bk*&bk*& bk \\ \hline
\end{tabular}
\]
}

{\tiny
\[
\setlength{\tabcolsep}{1 mm}
\begin{tabular}{|l|l|l|l|l|l|l|l|l|l|l|l|l|l|l|l|l|l|l|l|l|l|l|}
\hline
$n$ &165&168&171&174&177&180&183&186&189&192&195&198&201&204 \\ \hline
$k$ & 55& 56& 57& 58& 59& 60& 61& 62& 63& 64& 65& 66& 67& 68 \\ \hline
$d$ & 36& 36& 36& 36& 36& 38& 38& 38& 40& 41& 42& 42& 42& 41 \\ \hline
code&bk*&bk*&bk*&bk*&bk*& bk&bk*&bk*& bk& bk& bk&bk*&bk*& bk \\ \hline
\end{tabular}
\]
}

{\tiny
\[
\setlength{\tabcolsep}{1 mm}
\begin{tabular}{|l|l|l|l|l|l|l|l|l|l|l|l|l|l|l|l|l|l|l|l|l|l|l|}
\hline
$n$ &207&210&213&216&219&222&225&228&231&234&237&240&243&246 \\ \hline
$k$ & 69& 70& 71& 72& 73& 74& 75& 76& 77& 78& 79& 80& 81& 82 \\ \hline
$d$ & 43& 44& 44& 44& 45& 47& 48& 47& 46& 48& 48& 48& 49& 51 \\ \hline
code& bk& bk&bk*&bk*& bk& bk&bk*& bk&bk*& bk&bk*&bk*& bk& bk \\ \hline
\end{tabular}
\]
}

{\tiny
\[
\setlength{\tabcolsep}{1 mm}
\begin{tabular}{|l|l|l|l|l|l|l|l|l|l|l|l|l|l|l|l|l|l|l|l|l|l|l|}
\hline
$n$ &249&252&255 \\ \hline
$k$ & 83& 84& 85 \\ \hline
$d$ & 52& 53& 54 \\ \hline
code&bk*& bk&bk* \\ \hline
\end{tabular}
\]
}
\subsection{$4$-CIS codes}

For $1 \le k\le  \lfloor 256/t \rfloor$ except for $k=37$, we have checked that there are $4$-CIS $[tk, k]$ codes that are either ${\mbox{bk}}$ or ${\mbox{bk}}^*$. We have checked that the best known linear $[148, 37, 41]$ code in the Magma database~\cite{magma} is not 4-CIS. See the below tables.

{\tiny
\[
\setlength{\tabcolsep}{1 mm}
\begin{tabular}{|l|l|l|l|l|l|l|l|l|l|l|l|l|l|l|l|l|l|l|l|l|l|l|}
 \hline
$n  $&  8 & 12& 16& 20 & 24& 28&32& 36& 40& 44& 48& 52& 56& 60& 64& 68& 72\\ \hline
 $k  $&  2&  3& 4 & 5  &  6& 7 & 8&  9& 10& 11& 12& 13& 14& 15& 16& 17& 18\\ \hline
$d$  & 5  &6  &8  &9   &10 &12 &13& 14& 16& 16& 17& 19& 20&  21& 24  &24  &24\\ \hline
code & bk& bk*& bk*& bk& bk*&bk&bk& bk& bk& bk*&bk& bk&bk*& bk& bk* & bk* &bk\\
\hline
\end{tabular}
\]
}
{\tiny
\[
\setlength{\tabcolsep}{1 mm}
\begin{tabular}{|l|l|l|l|l|l|l|l|l|l|l|l|l|l|l|l|l|l|l|l|l|l|l|}
 \hline
$n  $& 76  &80 &84 &88 &92 &96 &100&104 &108 &112 &116 &120 &124 &128 &132 &136 & 140\\ \hline
 $k  $&19  &20 &21 &22 &23 &24 &25 &26  &27  &28  &29 &30   &31 &32   &33 &34 &35 \\ \hline
$d$  & 24  &25 &27 &28 &28 &28 &31 &32  &32  &32  &34 &34   &36 &36   & 37  &38  &40\\ \hline
code & bk  & bk&bk &bk &bk &bk &bk &bk  & bk*& bk*&bk &bk   &bk*&bk*  & bk & bk &bk*\\
\hline
\end{tabular}
\]
}
{\tiny
\[
\setlength{\tabcolsep}{1 mm}
\begin{tabular}{|l|l|l|l|l|l|l|l|l|l|l|l|l|l|l|l|l|l|l|l|}
 \hline
$n  $& 144  &148 &152 &156 &160 &164 &168&172 &176 &180 &184 &188 &192 &196\\ \hline
 $k  $&36  &37 &38 &39 &40 &41 &42 &43  &44  &45  &46 &47   &48 &49   \\ \hline
$d$  & 42  &? &40 &42 &44 &44 &44 &45  &46  &46  &48 &48   &50 &50  \\ \hline
code & bk  &? &bk* &bk &bk &bk*&bk*&bk  &bk& bk*  &bk &bk   &bk*&bk   \\
\hline
\end{tabular}
\]
}

{\tiny
\[
\setlength{\tabcolsep}{1 mm}
\begin{tabular}{|l|l|l|l|l|l|l|l|l|l|l|l|l|l|l|l|l|}
 \hline
$n  $ &200 &204 & 208  &212 &216 &220 &224 &228 &232&236 &240 &244 &248 &252 &256\\ \hline
 $k $ &50 &51 &52 & 53  &54 &55 &56 &57 &58 &59 &60  &61  &62  &63 &64       \\ \hline
$d$   &50 &52 &52 & 54  &52 &54 &55 &57 &60 &60 &58  &60  &62  &62 &64     \\ \hline
code &bk* &bk &bk*& bk* &bk*&bk*&bk &bk &bk &bk*&bk* &bk  &bk* &bk*&bk*    \\
\hline
\end{tabular}
\]
}

\subsection{$t$-CIS codes with $5 \le t \le 256$}

For $5 \le t \le 256$ and $1 \le k\le  \lfloor 256/t \rfloor$, all the best known codes
in the Magma database have been checked. We conclude that there are $t$-CIS $[tk,k]$ codes that are either ${\mbox{bk}}$ or ${\mbox{bk}}^*$.

\bigskip

\section{Construction Methods}
\label{sec-cstr}

For background on quasi-cyclic codes we refer the reader to \cite{LS1,LS2,LS3}.
In what follows, we give efficient ways to construct $t$-CIS codes in the sense that we can generate many $t$-CIS codes quickly and hence we easily get $t$-CIS codes with high minimum distances from them.

\begin{prop}[quasi-cyclic codes]
Let $C$ be a quasi-cyclic $[tk,k]$ code of co-index $k.$ Assume that $C$ is a 1-generator quasi-cyclic code with generating row
$$(a_1,a_2,\cdots,a_t),$$

where $a_i \in \F_2[x]$ are polynomials coprime with $x^k-1.$ Then $C$ is a $t$-CIS code.
\end{prop}

\begin{preuve}
The determinant of a circulant matrix of attached polynomial $a(x)$ is zero if and only if $deg(\gcd(a(x), x^k+1))>0.$ The result follows.
\end{preuve}

\begin{ex} The best known linear code $[243,9,118]$ is a 1-generator quasi-cyclic code with generating row in octal
$$[ 175, 177, 63, 357, 257, 253, 25, 73, 267, 113, 135, 377, 123, $$ $$ 337, 75, 37,
273, 51, 155, 153, 45, 35, 5, 65, 127, 133, 147 ].$$
The polynomials in $x$ corresponding to these $27$ numbers can be shown to be all coprime with $x^{27}-1.$
\end{ex}

The following constructions are a natural extension of those in \cite{2CIS, Kim01}.

\begin{lem}[Subtracting construction]\label{lem_subt}
Suppose that $C$ is a  $t$-CIS $[tk, k]$ code with generator matrix $G=(A_1~ A_2 ~ \cdots A_t)$, where each $A_j$ ($1 \le j \le t$) is an invertible $k \times k$ matrix. Then, there exists a $t$-CIS $[(k-1)t, k-1]$ code with generator matrix $G'=(A_1'~ A_2' ~ \cdots A_t')$, where each $A_j'$  ($1 \le j \le t$) is a $(k-1) \times (k-1)$ invertible matrix.
\end{lem}

\begin{preuve}
Choose any $i$ ($1 \le i \le k$). Delete the $i$th row of $G$. Then each $A_j$ ($1 \le j \le t$) becomes a $(k-1) \times k$ matrix whose rank is $k-1$. For each $j$ ($1 \le j \le t$), there exists a column of $A_j$ which is a linear combination of the rest columns of $A_j$.  Delete the column of $A_j$ to get $A_j'$, which is invertible since rank$(A_j')=k-1$.
\end{preuve}

\begin{prop}[Building up construction]\label{prop_building}
Suppose that $C$ is a $t$-CIS $[tk, k]$ code $C$ with generator matrix
$G=(A_1~ A_2 ~ \cdots A_t)$, where each $A_j$ ($1 \le j \le t$) is an invertible $k \times k$ matrix.
Let $A_j({\bf r}_i)$ ($1 \le i \le k$) denote the $i$th row of the matrix $A_j$.
Then for any vectors ${\bf x}_j \in \mathbb F_2^n$ ($1 \le j \le t$) and $y_{ij} \in \mathbb F_2$ ($1 \le i \le k, 1 \le j \le t$),
the following matrix $G_1$ generates a $t$-CIS $[t(k+1), k+1]$ code $C_1$.

\begin{equation}\label{eq:G_1}
 G_1 =
 \left(
   \begin{array}{c|c|c|c|c |c|c }
    z_1 & {\bf x}_1  & z_2 & {\bf x}_2 & \cdots &  z_t & {\bf x}_t   \\
 \hline
   y_{11} & A_1({\bf r}_1) & y_{12} & A_2({\bf r}_1) & \cdots & y_{1t} &  A_t({\bf r}_1) \\
       y_{21} & A_1({\bf r}_2) & y_{22} & A_2({\bf r}_2) & \cdots & y_{2t} &  A_t({\bf r}_2) \\
         \vdots & \vdots & \vdots & \vdots & \vdots & \vdots & \vdots \\
     y_{k1} & A_1({\bf r}_k) & y_{k2} & A_2({\bf r}_k) & \cdots & y_{kt} &  A_t({\bf r}_k) \\
      \end{array}
   \right)
      \end{equation}

   where for each $j$ ($1 \le j  \le t$), ${\bf x}_j$ satisfies
   ${\bf x}_j= \sum_{i=1}^k c_{ij} A_j({\bf r}_i)$ for uniquely determined $c_{ij}$'s ($c_{ij}=0,1$) and $z_j$ satisfies $z_j=1+ \sum_{i=1}^k c_{ij}  y_{ij}$.
\end{prop}

\begin{preuve} It is shown~\cite{2CIS} that the $(k+1)\times (k+1)$ matrix with the rows
$$z_1|{\bf x}_1, ~y_{11}| A_1({\bf r}_1),~ y_{21}| A_1({\bf r}_2), ~\dots,~ y_{k1}| A_1({\bf r}_k)$$
is invertible. By the same argument, we see that for any $1 \le j \le t$, the $(k+1)\times (k+1)$ matrix with the rows
$$z_j|{\bf x}_1, ~y_{1j}| A_j({\bf r}_1),~ y_{2j}| A_j({\bf r}_2), ~\dots,~ y_{kj}| A_j({\bf r}_k)$$
is invertible.
Therefore, $G_1$ is a $[t(k+1), k+1]$ $t$-CIS code.
\end{preuve}

\begin{prop}\label{prop:converse} Let $C$ be a $t$-CIS $[tk, k]$ code $C$. Then it is equivalent to a $t$-CIS $[tk, k]$ code $C_1$ which is constructed from a  $t$-CIS $[t(k-1), k-1]$ code by using Proposition~\ref{prop_building}.
\end{prop}

\begin{preuve}
The key idea of this proof is given in the proof of Proposition VI.6~\cite{2CIS}. Basically, the subtracting construction (Lemma~\ref{lem_subt}) and the building up construction (Proposition~\ref{prop_building}) are reversible operations.
\end{preuve}

\medskip

Similar to the mass formula for $2$-CIS codes given in Prop. VI. 9 in~\cite{2CIS}, there exists a formula for determining if a list of $t$-CIS codes is complete.

\begin{prop}
For positive integers $k$ and $t \geq 2$, let ${\bf C}$ denote the set of all $t$-CIS $[tk,k]$ codes and let $S_{tk}$ the symmetric group of degree $tk$ act on the columns of elements of ${\bf C}$.  Suppose $C_1, \dots, C_s$ are representatives from each equivalence class in ${\bf C}$ under the action of $S_{tk}$.  Let ${\bf C}_{sys}$ denote the set of all $t$-CIS $[tk,k]$ codes with generator matrix in the form $(I_k | A_1 | \cdots |A_{t-1})$ with all $A_j \in GL(k,2)$.  Assume each $C_i$ is in ${\bf C}_{sys}$.  Then
\begin{equation}\label{tCIS_mass}
{g_k}^{t-1} = \sum_{i=1}^s |Orb_{S_{tk}}(C_i) \cap {\bf C}_{sys} |
\end{equation}
in which $Orb_{S_{tk}}$ denotes the orbit of $C_i$ under $S_{tk}$ and $g_k$ denotes the cardinality of $GL(k,2)$. Hence we omit the details.
\end{prop}

\begin{preuve}
The proof of this proposition is just the repetition of the argument in the proof of~\cite[Prop. VI.7]{2CIS}.
\end{preuve}

\begin{ex}

\end{ex} Let $C$ be a $[3,1,3]$ repetition code with generator matrix $(1~1 ~1)$, which is $3$-CIS.
 Choose ${\bf x}_1 = (0), {\bf x}_2=(1), {\bf x}_3=(0), y_{11}=0, y_{12}=0,$ and $y_{13}=1$. Then $z_1=1, z_2=1$ and $z_3=1$ by Proposition~\ref{prop_building}. Hence the following matrix

\[
G_1 =
 \left(
   \begin{array}{c|c|c|c|c|c}
   1 & 0 & 1 & 1 & 1& 0 \\
   \hline
   0 & 1 & 0 & 1 & 1& 1 \\
   \end{array}
   \right)
   \]
   generates a $[6,2,4]$ $3$-CIS code.

\section{On the Classification of $t$-CIS Codes}
\label{sec:class-CIS}

In this section, we describe two methods of classification for $t$-CIS codes relating to the equivalence classes of matrices under defined equivalence relations.

\subsection{Classification Methods Using Equivalence Classes of Matrices}


The two methods we propose here are based on the classification method given in~\cite{frei}.  For the first method we consider the following notion of equivalence on $GL(k, 2)$: two matrices $A,B \in GL(k, 2)$ are equivalent, $A \sim_1 B$, if and only if $A = BP$ where $P$ is a $k \times k$ permutation matrix.  Let $[GL(k,2)]^{t-1}$ denote the set of all $k \times k(t-1)$ concatenations of $t-1$ elements from $GL(k,2)$.   For the second method we consider a notion of equivalence on $[GL(k,2)]^{t-1}$:   two matrices $A,B \in [GL(k,2)]^{t-1}$ are equivalent, $A \sim_2 B$, if and only if $A = P_k B P_{k(t-1)}$ where $P_k$ is a $k \times k$ permutation matrix and $ P_{k(t-1)}$ is a $k(t-1) \times k(t-1)$ permutation matrix.  Note that the method applied in~\cite{frei} is the case where $t=2$.

\begin{rem}
$\sim_1$ and $\sim_2$ are equivalence relations.
\end{rem}

\begin{prop}
(Method 1)  Given a set $S_{t-1}$ of all representatives of inequivalent $(t-1)$-CIS codes of dimension $k$ a set of all inequivalent $t$-CIS codes, $S_{t}$, of dimension $k$ is obtained by
\begin{enumerate}
\item Appending a matrix representative of each equivalence class under $\sim_1$ to each code to obtain a set $\overline{S_{t}}$ of $t$-CIS codes.
\item Carrying out an equivalence check on $\overline{S_{t}}$ to eliminate equivalent codes, then keeping only one representative from each class we obtain $S_t$.
\end{enumerate}
\end{prop}

\begin{prop}
(Method 2)  Given a set $Cat_{k,t-1}$ of all representatives of inequivalent (under $\sim_2$) $k$ by $k(t-1)$ matrices in $[GL(k,2)]^{t-1}$ a set of all inequivalent $t$-CIS codes, $S_{t}$, of dimension $k$ is obtained by
\begin{enumerate}
\item Appending a matrix representative in $Cat_{k,t-1}$ to the identity matrix $I_k$ to form a set  $\overline{S_{t}}$ of $t$-CIS codes.
\item Carrying out an equivalence check on $\overline{S_{t}}$ to eliminate equivalent codes, then keeping only one representative from each class we obtain $S_t$.
\end{enumerate}
\end{prop}

Previously in~\cite{frei}, the cardinality of  $Cat_{k,t-1}$ is given for $t=2$ and $k=1,2,3,4,5,6,7$.  In the following proposition we extend to some results for $t=3$.  The values are found using graph isomorphism and all code equivalences are also checked using graph isomorphism as described in~\cite{frei}.  All classifications were implemented in Magma~\cite{magma}.

\begin{prop}
The cardinality of $Cat_{k,t-1}$ for $t=3$ and $k=1,2,3,4$ is given in the following table:
{
\[
\begin{array}{|r|c|c|c|c|}
\hline
$k $ & 1 & 2 & 3 & 4  \\
\hline
\text{Total} & 1 & 4 & 58 & 4822  \\
\hline
\end{array}
\]
}
\end{prop}

In the following we use the methods described above to obtain a classification of $3$-CIS codes for $k=1,2,3,4$.

\subsection{Classification of short $t$-CIS codes}

\begin{rem}
We note that any $t$-CIS code has minimum distance $\geq t$.
 If not then there exists at least one information set which has an all zero row in the corresponding column submatrix (this is a contradiction since the rank of the submatrix must be full).
\end{rem}

The following table gives a summary of the classification.  This classification was obtained using the methods 1 and 2 described above.  The $i$th column $(i=2$--$5)$ gives the number of $3$-CIS codes with $d=i+1$, and in the parenthesis are the number of which are self-orthogonal and then not self-orthogonal.  The final column gives a sum total of $3$-CIS codes corresponding to the length in column 1.

{\tiny
\[
\setlength{\tabcolsep}{0.5 mm}
\begin{array}{l|l|l|l|l|l|l}
\hline
3k & d =3 & d=4 & d=5 & d=6 & d=7 & \text{Total \#}\\
\hline
3 & 1 (0+1) & & & & &  1\\
6 & 2 (0+2) &1 (1+0) & & & & 3  \\
9 & 11 (0+11) &8 (1+7) & & & & 19  \\
12 & 170 (0+170) & 178 (6+ 172) &12 (0 +12) & 1 (0+1) & &361 \\
15 & 10904  & 15842  & 2543  & 91 & 1  & 29372 \\  
   & (0+10904)& (15+15827) & (0+2543)& (1+ 90)  &(0+1) \\  
\hline
\end{array}
\]
}

We summarize our classification as follows.

\begin{thm}
For each $3k \in \{3,6,12,15 \}$, there is a unique optimal 3-CIS code of length $3k$.
There are eight optimal 3-CIS $[9,3,4]$ codes.
\end{thm}

\section{Conclusion and Open Problems}
\label{sec-concusion}

In this paper we have introduced and studied $t$-CIS codes for $t>2.$ The main tool is an algorithm for CISness testing based on the Edmonds base partitioning algorithm from matroid theory. Combining this algorithm with the BKLC function of Magma \cite{magma}, which itself is based on Grassl's tables \cite{Grassl}, we were able to show that
for each $t$ and $k$ such that $3 \le t \le 256$ and $1 \le k\le  \lfloor 256/t \rfloor$ except two cases, there are $t$-CIS codes of dimension $k$ that are optimal or with best known parameters. The only open pairs of parameters where this approach fails so far
 are $t=3$ with $k=44$ and $t=4$ with $k=37$. The approaches of~\cite{2CIS} have also been visited in turn. The $\Z_4$-codes have been used successfully to create Boolean functions with better correlation immunity than linear ones. A recently discovered $\Z_4$-code of parameters
$(24,4^6,18)$ has been applied. It would be interesting to know if it can be completed into an infinite family.
 In the asymptotic domain, the existence of long $3$-CIS codes that are good has been proved. It remains to be seen if there are families of good long binary codes of rate $1/t$ that are not $t$-CIS. Are almost all codes of rate $1/t$ $t$-CIS on average? We do not know the answer to this question.

\section*{Acknowledgement}

 P. Sol\'{e} thanks Sogang University for its warm hospitality and also thanks Neil Robertson for helpful discussions. We thank the reviewers for their constructive comments on our paper.

\bigskip

\newpage 

\bibliography{sca}
\bibliographystyle{IEEE}

{\bf Claude Carlet}~~
Claude Carlet received the Ph.D. degree from the University of Paris 6, Paris, France, in 1990 and the Habilitation to Direct theses from the University of Amiens, France, in 1994.
He was with the Department of Computer Science at the University of Amiens from 1990 to 1994 and with the Department of Computer Science at the University of Caen, France, from 1994 to 2000. He is currently Professor of Mathematics at the University of Paris 8. His research interests include coding theory, Boolean functions and cryptology.
Professor Carlet was Associate Editor for Coding Theory of IEEE Transactions on Information Theory from March 2002 until February 2005. He is the Editor in Chief of the journal ``Cryptography and Communications - Discrete Structures, Boolean Functions and Sequences" (CCDS) published by SPRINGER. He is in the editorial boards of the journals ``Designs, Codes and Cryptography" (SPRINGER), ``International Journal of Computer Mathematics" (Taylor \& Francis) and ``International Journal of Information and Coding Theory" (Inderscience publishers).

\medskip

{\bf Finley Freibert}~~
Finley Freibert received a B.A. in mathematics from DePauw University, USA in 2006 and a Ph.D. degree in Applied and Industrial Mathematics from University of Louisville, KY, USA in 2012. He was an Assistant Professor at the Department of Mathematics, Ohio Dominican University, Columbus, USA. His research interests include algebraic coding theory.

\medskip

{\bf Sylvain Guilley}~~
Sylvain Guilley is professor at TELECOM-ParisTech. His group works on
the proven security of electronic circuits and embedded systems. His own
research interests are trusted computing, cyber-security, secure
prototyping in FPGA and ASIC, and formal methods. Sylvain has authored
more than one hundred research papers, and about ten patents. He is
member of the IACR, the IEEE and senior member of the CryptArchi club.
He is alumni from Ecole Polytechnique and TELECOM-ParisTech. In 2010, he
has co-founded the Secure-IC company as a spin-off of TELECOM-ParisTech.
Since 2012, he organizes the PROOFS workshop, which brings together
researchers whose objective is to increase the trust in the security of
embedded systems.

\medskip

{\bf Michael Kiermaier}~~
Michael Kiermaier received the Diploma degree in mathematics from the
Technical University of Munich, Germany, in 2006 and the Ph.D. degree
from the University of Bayreuth, Germany, in 2012. His research
interests include coding theory, design theory and finite geometry.

\medskip

{\bf Jon-Lark Kim}~~
Jon-Lark Kim (S'01-A'03) received the B.S.degree in mathematics from POSTECH, Pohang, Korea, in 1993, the M.S. degree in mathematics from Seoul National University, Seoul, Korea, in 1997, and the Ph.D. degree in mathematics from the University of Illinois at Chicago, in 2002. From 2002 to 2005, he was with the Department of Mathematics at the University of Nebraska-Lincoln as a Research Assistant Professor. From 2005 to 2012 he was with the Department of Mathematics at the University of Louisville, KY as an Assistant Professor and an Associate Professor. Currently, he is an Associate Professor at the Department of Mathematics, Sogang University, Korea from 2012. He was awarded a 2004 Kirkman Medal of the Institute of Combinatorics and its Applications. He is a member of the Editorial Board of both ``Designs, Codes, and Cryptography'' (2011-current) and ``International J. of Inform. and Coding Theory'' (2009-2013). His areas of interest include Coding Theory and its interaction with Algebra, Combinatorics, Number Theory, Cryptography, and Industrial Mathematics.

\medskip

{\bf Patrick Sol\'{e}}~~
Patrick Sol\'{e} received the Ing\'{e}nieur and the Docteur Ing\'{e}nieur degrees fromEcole Nationale Sup\'{e}rieure des T\'{e}l\'{e}communications, Paris, France in 1984 and
1987, respectively, and the Habilitation à Diriger des Recherches degree from
Universit\'{e} de Nice, Sophia-Antipolis, France, in 1993.
He has held visiting positions at Syracuse University, Syracuse, NY, during
1987–1989, Macquarie University, Sydney, Australia, during 1994–1996, and at
Universit\'{e} des Sciences et Techniques de Lille, Lille, France, during 1999–2000.
He has been a permanent member of Centre National de la Recherche Scientifique
since 1989, and with the rank of Research Professor (Directeur de
Recherche) since 1996. Since 2011, he has a joint affiliation with King AbdulAziz
University, Jeddah, Saudi Arabia.
He was associate editor of the Transactions from 1999 until 2001. He is currently
associate editor of Advanced in Math of Communication.
His research interests include coding theory (covering radius, codes over
rings, geometric codes, quantum codes), interconnection networks (graph
spectra, expanders), space time codes (lattices, theta series), and cryptography
(Boolean functions). He is well-known for using methods of ring theory in
algebraic coding. In particular the best paper award in information theory was
awarded in 1994 to his paper on codes over the integers modulo 4, which has
been quoted 720 times since. He is the author of more than a hundred and
twenty Journal papers, and one book (Codes Over Rings, World Scientific,
2008).



\end{document}